# Highly Reproducible and CMOS-compatible VO$_2$-based Oscillators for Brain-inspired Computing

*Olivier Maher\*, Roy Bernini, Nele Harnack, Bernd Gotsmann, Marilyne Sousa, Valeria Bragaglia, and Siegfried Karg\**

IBM Research Zurich, Säumerstrasse 4, 8803 Rüschlikon, Zürich, Switzerland

E-mail: OGM@zurich.ibm.com, Roy.Bernini@hotmail.com, NEH@zurich.ibm.com, BGO@zurich.ibm.com, VBR@zurich.ibm.com and SFK@zurich.ibm.com

**ABSTRACT**. With remarkable electrical and optical switching properties induced at low power and near room temperature (68 ˚C), vanadium dioxide (VO$_2$) has sparked rising interest in unconventional computing among the phase-change materials research community[1]. The scalability and the potential to compute beyond the von Neumann model make VO$_2$ especially appealing for implementation in oscillating neural networks for artificial intelligence (AI) applications, to solve constraint satisfaction problems, and for pattern recognition[2–5]. Its integration into large networks of oscillators on a Silicon platform still poses challenges associated with the stabilization in the correct oxidation state and the ability to fabricate a structure with predictable electrical behavior showing very low variability[6,7]. In this work, the role played by the different annealing parameters applied by three methods (slow thermal annealing, flash annealing, and rapid thermal annealing), following the vanadium oxide atomic layer deposition (ALD), on the formation of $VO_2$ grains is studied and an optimal substrate stack configuration that minimizes variability between devices is proposed. Material and electrical characterizations are performed on the different films and a step-by-step recipe to build reproducible $VO_2$-based oscillators is presented, which is argued to be made possible thanks to the introduction of a hafnium oxide ($HfO_2$) layer between the silicon substrate and the vanadium oxide layer. Up to seven nearly identical $VO_2$-based devices are contacted simultaneously to create a network of oscillators, paving the way for large-scale implementation of $VO_2$ oscillating neural networks.

(INTRODUCTION)

With phenomenal computing and learning capabilities far beyond the fastest chips, the brain remains today the most power-efficient computational system. Implementing brain-like circuitry for faster less power-hungry massive data processing drives the industry to downscale CMOS technology and innovate towards new 'neurmorphic' materials. Research surrounding phase-change materials has demonstrated their ability to mimic some of the brain's elemental operations, both synaptic and neuronal. For example, the most recent work on $Ge_2Sb_2Te_5$ memories enables vector-matrix multiplication for in-memory computing, implemented in a spiking neural network (SNN)[8]. This type of phase-change spiking system is ideal both for powering biologically-realistic AI applications and for its technological potential in terms of integration at the back-end-of-the-line. Recently, an analogous computing technique based on another phase-change material system offering comparable benefits has sparked a growing interest among several research groups[2,9,10]. The ambition is to unravel further the potential of phase-change materials by using oscillation-based computing, inspired by the rhythmic patterns of action potentials exhibited by neurons during the learning phase[11]. This approach provides new computational paradigms for bio-inspired applications.

In an oscillating neural network (ONN), the information is carried in the phase relations between coupled oscillators rather than in the amplitude of the signals, making them intrinsically resistant to voltage-scaled noise and typical input pattern distortion problems in Machine Learning (ML)[6]. This method has demonstrated the potential to perform all types of arithmetic computation and shows the promise of major computational improvements, especially for optimization tasks, Boolean satisfiability (3-SAT) problems, and Ising machine problems[5,9–12]. Typically, finding the best solution to optimization tasks in traditional computers demands a tremendous amount of power and far larger computing times. A lightweight hardware design with phase-change material electronics offering moderate accuracy performed locally in-memory is often sufficient to find a suitable solution to common industrial operations[10]. Additionally, computing with ONNs avoids the von Neumann bottleneck energy costs arising from data transfer from the memory unit to the processor by embracing the fundamental principle of in-memory computation: "let physics do the computing"[10,12,13]. Consequently, an ONN architecture built with nanoscale oscillators and compact high-fanout interconnections should be favored to allow for a richer representation of information while avoiding long-range power-hungry coupling between devices[2,7,14].

Assembling an ONN depends on the successful fabrication of individual oscillators with predictable and reproducible behavior. An ideal oscillator needs to be scalable, offering easy integration on a Silicon (Si) platform, power-efficient, endurant, and operable at high frequency. Niobium oxide[15] ($NbO_2$), magnesium-oxide-based magnetic tunnel junction[16] (MgO-based STO), and vanadium dioxide[1] ($VO_2$) all exhibit oscillatory capabilities, with the latter being the leading candidate thanks to the low power required to trigger a high-frequency resistive switching near room temperature (68 °C)[1,17]. The source mechanisms behind $VO_2$'s switching behavior are still being debated in the research community. For most studies, it originates between a Peierls and a Mott phase transition[18]. This little understanding on $VO_2$'s intrinsic phase-transition nature as well as the complexity required to connect several devices for the integration of a large-scale ONN has limited most of the research to simulation-based results or generally, to the study of only two coupled $VO_2$-based oscillators[17,19,20]. In addition, the high variability between $VO_2$-based oscillators, originating from the multitude of metastable vanadium oxide oxidation states generated during fabrication, has been reported to be a main limiting factor for advancing full hardware-based ONN approaches. For example, in Won *et al.*[21] and Pósa *et al.*[22], it is demonstrated that even the highly controllable magnetron sputtering deposition technique encounters challenges in growing pure $VO_2$ on a silicon dioxide ($SiO_2$) substrate. In fact, the variable $VO_2$ transition temperatures obtained in Pósa *et al.*[22] make devices fabricated by magnetron sputtering difficult to scale to nanodimensions and unfit for the large circuit implementation we are targeting. Other studies[3,4,9,23,24] also established that the fabrication of $VO_2$ layers on amorphous substrates through pulsed laser deposition, chemical vapor deposition, sputtering, and ALD tends to result in polycrystalline films with granular structure and considerable surface roughness[25]. The fabrication challenges associated with $VO_2$ deposition (magnetron sputtering or PLD-deposition[22]) include the coexistence of several compositions, such as the Magneli ($V_nO_{2n-1}$, where $3 \leq n \leq 9$) and the Wadsley ($V_nO_{2n+1}$, where n = 1–6) phases that are all – including $VO_2$ – intermediary phases towards the most thermodynamically stable stoichiometry of $V_2O_5$ [22,26]. Additionally, achieving precise control over the oxidation state, topography, crystal orientation, and degree of crystallinity in $VO_2$ crystals poses challenges in fabricating devices with high performance yields[27–29]. These morphology challenges have been shown to contribute to the undesired variability among electrical VO$_2$-based oscillators[30].

In an effort to mitigate this variability and produce CMOS process-compatible vanadium-oxide oscillators to build ONNs, we study the advantages of several annealing techniques post-ALD deposition on different stacks where we fine-tune their respective parameters[4]. We focus our analysis on the nanoscale granular morphology of the layer, as previous work[6,30] has demonstrated that it has the greatest impact on the transition characteristics and performance of our developed VO$_2$ devices. We also look into inhomogeneities resulting from variations in chemical composition and the coexistence of other oxidation states of vanadium oxide, as they are expected to contribute to device variability[26]. In this work, we investigate the influence of the annealing parameters on the VO$_2$'s nanoscale structure and composition through Atomic Force Microscopy (AFM), Raman Spectroscopy, and X-ray reflectometry (XRR), as well as the VO$_2$'s resulting electrically prompted crystalline phase transition. Our goal is to obtain a granular film of high quality, which we define as a dense layer of compact grains with smooth surface roughness, uniform and reproducible electrical behavior showing a sharp and narrow resistance-temperature (R-T) hysteresis[23]. Finally, we also exploit the properties of different metal-oxides to study, influence, and engineer the crystalline phase-change temperature, opening the door for wider industrial applications where the thermal budget caused by the heat of the operating peripheral circuitry is limited[31]. We show how the results obtained can be extended to various device topologies where layers are staked between electrodes to realize a network of VO$_2$ oscillators.

**METHODS**

**Vanadium Oxide Film Deposition**

Vanadium oxide can be synthesized through numerous methods, including gas-phase techniques like pulsed laser deposition (PLD), chemical vapor deposition (CVD), sputtering, and atomic layer deposition (ALD) as well as solution-based methods, such as sol-gel processes and hydrothermal synthesis. While magnetron sputtering is a widely-used deposition technique, the high energy of the deposited particles leads to limited results for vanadium oxide films below 100 nm [32]. To achieve scalable ultrathin films suitable for low-power applications, gas-phase techniques offer better results[32]. [26] These methods are well-suited for forming thin layers directly on a heated substrate in the presence of an appropriate process gas or in a two-step process with an annealing step following the deposition.

We use ALD for our depositions, as it provides uniform growth on high aspect ratio nanostructures, homogeneous deposition over a large area, and CMOS compatibility with the possibility to co-integrate the grown material on top of underlying pre-existing circuitry[29,33]. We apply the novel Tetrakis[ethylmethylamino] vanadium (TEMAV) reaction on a Si platform, which calls for the presence of an inert carrier gas (Argon) and an oxidation agent (water) to coat uniformly and quickly[29]. We prefer a dense water-grown film over an ozone-grown layer, as it was shown to more likely give rise to $VO_2$ grains in the M1 crystallization phase after a post-annealing treatment[28,29]. Our process occurs at 150 °C to prevent the material (98% TEMAV) from being too volatile at low temperatures or from thermal decomposition at temperatures higher than 175 °C [29]. The in-house recipe deposition rate is about 0.5 Å per cycle (30-40 seconds), including a sequence of dosing-purging steps that keep the pressure in the chamber below 11 Pa. The final thickness of the layer is 60 nm, as confirmed by XRR measurements (Table 1).

After the ALD process, a post-deposition treatment at temperatures above 420 °C is necessary to transform the film into the desired oxidation state ($VO_2$)[29]. The ALD process alone provides a highly homogeneous vanadium-oxide layer but lacks sufficient control over crystallinity, stoichiometry, and phase required for the devices[29]. The necessary post-deposition annealing step typically degrades the film's morphology, and the effect is even stronger in thin films where the impact is directly measurable on the metal-to-insulator temperature (MIT) and on the properties of the $VO_2$ [34]. Our aim is to reduce the rough morphology obtained after annealing, which can be attributed to a volume change associated with the de-wetting process of the dielectric surfaces during annealing[29].

**Slow Thermal Annealing (STA)**

The Neocera Combinatorial 180 PLD system used for this method offers a precise control over the oxygen pressure with heating and cooling temperature ramps limited to a maximum of 25 °C $min^{-1}$. The films were brought to a set temperature of 520 °C at the sample holder, corresponding to a temperature of about 420 °C in the sample, with an oxygen flow defined to keep the pressure at 5 Pa at the highest temperature setpoint. Increasing the temperature beyond this point induces the crystallization of large grains and can raise the final oxidation state[29]. Upon reaching the set temperature of 520 °C, the samples were kept in the heating chamber for 5 to 10 minutes, before being cooled back down to room temperature.

**Flash Annealing (FLA)**

High heating and cooling rates during the annealing process offer increased control over the grains' nanostructure[35]. In the early stages of phase formation, nucleation occurs, and a rapid heating rate within the sample affects both grain growth and density, resulting in a smoother film with small compact structures[35]. Other experiments have also shown that a flash annealing step could improve the film smoothness and transition sharpness[36].

Using the flash lamp FLA-50AS, Dresden Thinfilm Technology, we investigate the possibility of growing small grains with the energy of a flash only. The flash is 20 ms long, with a power between 90 and 110 J cm$^{-2}$. The samples were preheated with a thermally regulated Si carrier wafer at a fixed temperature, ranging between 140 °C and 330 °C. This tool provides heating and cooling rates orders of magnitude faster than the STA, where exact values can only be estimated. The tool allows for oxygen pressure to be constrained within the range of 1.33 Pa to 66.66 Pa.

**Rapid Thermal Annealing (RTA)**

To achieve control over the grains' structure through rapid thermal rates, while still maintaining atmospheric conditions suitable for VO$_2$ growth, we used the ANNEALSYS AS-Micro RTP-System to explore rapid thermal annealing[26,35]. The samples were placed in a chamber and stabilized to a temperature of 300 °C, before being rapidly brought up to the final annealing temperature. The oxygen partial pressure, between 5 Pa and 25 Pa, was maintained through a continuous flow, and annealing times varied from 30 s to 600 s. This last technique combines the benefits of the STA and Flash annealing techniques by offering ultralow partial pressures with a heating rate of 18 °C s$^{-1}$.

**Device Configuration**

A planar configuration (see Figure 2c, with corresponding scanning electron microscopy (SEM) image in Figure 1c) was chosen for the fabrication of rectangular test structures with active region dimensions varying between 400 × 400 × 60 nm$^3$ and 2000 × 2000 × 60 nm$^3$. The irregularity of the grains at the nanoscale can define a preferential current path in the device, as observed by X-Ray diffraction (XRD) nanoimaging in Shabalin *et al.*[24] The advantage of this device geometry consists in few processing steps, offering a top view of the grains involved in the phase transition. However, this implies that a different preferential current path is created in each device, which is the source of significant variability. These tests structures are convenient to verify the reliability of our annealing process and to perform material characterization with a fast turnaround. To achieve coupling between oscillators, the crossbar configuration shown in Figure 1d is preferable, as it confines the current path within the intersection of the cross-section between a top and a bottom electrode (see Figure 2c).

The thickness of the $VO_2$ layer defines a current path and the amount power required to induce the phase transition. A 50 nm thick bottom electrode (3 nm Titanium (Ti) / 47 nm Platinum (Pt)) was embedded into the substrate if it was at least 50 nm thick through an e-beam lithography step followed by a dry etch. In the case of a thin substrate ($SiO_2$ < 50nm), the bottom electrodes were deposited directly on the substrate.

It should be noted that the planar and crossbar device configurations lead to different performance results. For instance, the crossbar configuration excels in localizing the current path, but assessing the film quality where the phase-transition occurs remains challenging due to characterization limitations. Evaluation at larger scales non-invasively through XRD, XRR, or Raman spectroscopy techniques is necessary, and device performance is subsequently correlated with these measurements under reasonable assessments.

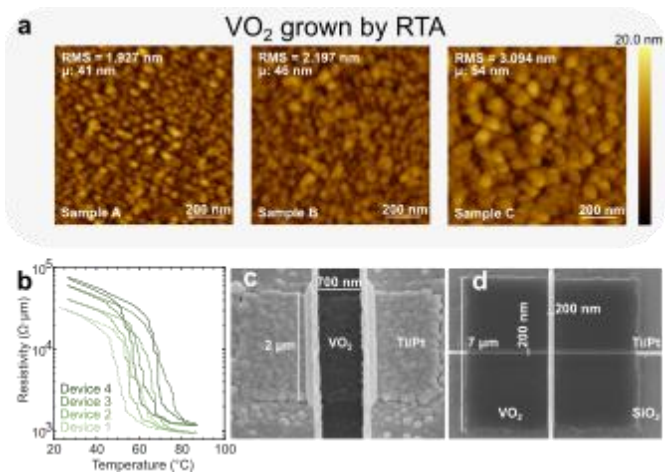

**Figure 1. VO$_2$ grown by STA. a** AFM measurements of an STA treated VO$_2$ film on a 1 μm thermal SiO$_2$ substrate grown with different annealing parameters. Sample A: 520 °C, 5 Pa O$_2$ pressure, 5min long annealing. Sample B: 520 °C, 5 Pa O$_2$ pressure, 10 min annealing time. Sample C: 540°C, 5 Pa O$_2$ pressure, 5 min annealing time. **b** R-T characteristics of VO$_2$ crossbar devices (active area: 200 nm × 200 nm × 60 nm) fabricated from Sample A. The variation in grain morphology and stoichiometry, as discussed in the following sections, causes the grains to switch at different temperatures, leading to jumps in the hysteresis curve. SEM images of a VO$_2$ planar device (active area: 2 μm × 700 nm × 60 nm) in **c** and a crossbar device (active cross-section area dimensions: 200 nm × 200 nm × 60 nm) in **d**.

## RESULTS

Different annealing techniques are explored to achieve high VO$_2$ granular quality. Synthesizing films in the VO$_2$ oxidation state only is limited by a fine window of homogeneity and a high sensitivity to the oxygen partial pressure during the thermal treatment[37]. These challenges have a direct impact on the reproducibility of the film's granular composition and surface roughness that affects the overall electrical performance[37]. A compact and dense layer of small grains favors low variability between devices[23].

Synthesizing this reproducible VO$_2$ layer with small and compact grains relies on precise control of annealing parameters, including temperature, time, heating/cooling rates, and oxygen partial pressure. Decoupling the effect of each parameter proves difficult due to the complex chemical and physical mechanisms happening during annealing. Therefore, we assess the overall impact of these parameters on the grains' morphology, oxidation phase, and surface roughness ex-situ and post-treatment through XRR, Raman, and AFM measurements.

We investigate the potential of three post-deposition annealing techniques to grow polycrystalline VO$_2$ for our oscillators. These techniques provide varying degrees of control over the annealing parameters.

### VO$_2$ Grown By Slow Thermal Annealing (STA)

The AFM micrographs shown in Figure 1a correspond to VO$_2$ samples *A*, *B*, and *C* grown directly on a 1 μm thick SiO$_2$ substrate. Various annealing treatments applied to the samples lead to nanoscale VO$_2$ grains of different dimensions. Samples *A*, *B*, and *C* were annealed at 520 °C, 520 °C, and 540 °C for 5 minutes, 10 minutes, and 5 minutes, respectively. Samples annealed at temperatures below 520 °C did not grow VO$_2$ grains. The conditions that lead to the smoothest film with the lowest surface roughness value (RMS) involve a short annealing time of 5 minutes at a set temperature of 520 °C, achieved with a ramp rate of 25 °C min$^{-1}$ under an oxygen partial pressure of 5 Pa. In addition to these conditions, our findings in Figure 1 reveal that VO$_2$ grows a rougher surface when there is an increase in annealing temperature or time, due to the formation of larger grains. This is consistent with similar results found in other studies[26,35,38].

Figure 1b shows the R-T characteristics of crossbar devices of identical dimensions (active area of 200 nm × 200 nm × 60 nm) fabricated from the smoothest film (sample *A*). The high variability among the samples is evident in both the jumps observed in the hysteresis curves, attributed to grains of various morphology switching at different temperatures, and the different resistivity values in the insulator- and metallic-states. The annealing conditions employed to fabricate these devices served as an initial reference for further optimization using alternative annealing methods.

### VO$_2$ Grown By Flash Annealing (FLA)

Figure 2a shows the Raman spectra of films annealed with a flash power of 90 J cm$^{-2}$ and an oxygen pressure of 20 Pa at chuck temperatures ranging from 185 °C to 250 °C. The results indicate that pre-heating the sample at least to 215 °C is necessary to measure the characteristic VO$_2$ Raman peaks at 193 cm$^{-1}$ and 223 cm$^{-1}$ [39]. As the chuck temperature is raised above 215 °C before flashing, the intensity of the Raman peaks grows, indicating a greater amount of the initially amorphous material successfully crystallizes into VO$_2$ grains. A lower ratio of the intensity of the main Raman peaks (193 cm$^{-1}$ and 223 cm$^{-1}$) relative to the intensity of the weakest peaks (389 cm$^{-1}$, 497 cm$^{-1}$, 612 cm$^{-1}$, …) in VO$_2$ is also associated with an increase in surface roughness in our samples[40]. The presence of V$_2$O$_5$ is detected (145 cm$^{-1}$) when the pre-heating temperature is 250 °C or higher.

In Figure 2a, the average grain diameter size (µ) and mean surface roughness (RMS) values extracted from AFM are plotted against the chuck temperature before flash annealing. The grain size remains stable around 25 nm across all temperatures, while the surface roughness consistently increases with the substrate temperature. These trends suggest that only the surface roughness is influenced by an increase in substrate temperature. This observation is also supported by the sharper peaks measured by Raman spectroscopy at higher chuck temperatures (see Figure 2a). Comparing results in Figure 1a to Figure 2a indicates that flash annealing reduces surface roughness by over 50% compared to STA-annealed samples.

Figure 2a also shows 4-probe measurements performed on samples with the highest crystallized material content (pre-flash temperature: 250 °C). For the other samples with chuck temperatures below 215 °C, no resistive switching behavior was measured, which is consistent with the Raman spectra in Figure 2a.

A change in resistivity of nearly 1.5 orders of magnitude is observed in Figure 2a, with a hysteresis width of 6 °C and an IMT at 58 °C. This temperature is considerably lower than the typical 68 °C for VO$_2$ prepared by STA. This difference may arise from the presence of lattice defects, resulting from various phenomena such as low-angle grain boundaries, twinning, interface dislocations, or surface roughening[41,42]. The strain induced by relaxation paths, often challenging to monitor, could also influence the transition behavior of the annealed layer[42]. Studies[41] have demonstrated that typical lattice defects in granular $VO_2$ tend to lower its transition temperature. Additionally, the large transition window observed in Figure 2a suggests that interactions between grains or between grains and the substrate, possibly through thermoelastic behavior, could elongate the phase-transition hysteresis and reduce the apparent transition temperature[41,43].

The R-T measurements in Figure 2a also reveal a substantial variation in the resistivities (both insulator- and metallic-state) among samples 1 and 2 flash annealed under the same conditions, indicating a low level of reproducibility for this annealing technique.

Although flash annealing significantly reduces surface roughness (AFM measurements in Figure 2a) compared to samples grown by STA (see Figure 1a), the variability in grain dimensions, oxidation states, transition temperature, and electrical response between identically treated samples renders flash annealing unsuitable for the fabrication of uniform $VO_2$-based oscillators.

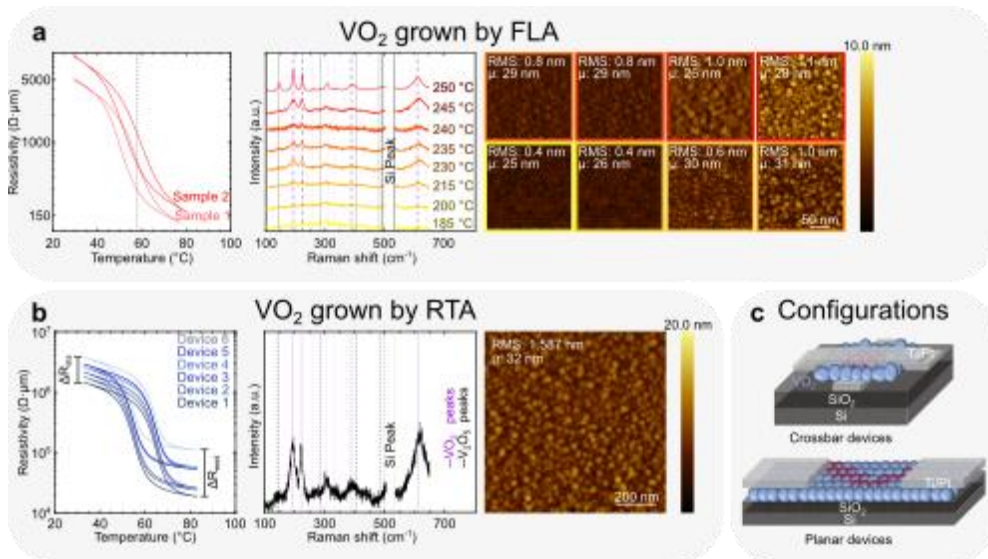

**Figure 2. VO$_2$ grown by FLA and RTA. a** R-T characteristics of samples FLA annealed with the conditions: Pre-flash temp: 250 °C, 26.66 Pa O$_2$ pressure, 20 ms long annealing:, Flash power: 90 J cm$^{-2}$. The transition of 1.5 orders of magnitude around 58 °C is highlighted with a dashed line. Raman spectra of vanadium-oxide samples grown on 1 μm thermal SiO$_2$ pre-heated at different temperatures before being flashed: Pre-flash temp: 185-250 °C, 26.66 Pa O$_2$, 20 ms long annealing, Flash power: 90 J cm$^{-2}$. Minimum substrate temperature required to anneal amorphous VO$_x$ into VO$_2$: 215 °C. AFM measurements of the samples pre-heated at different temperatures. **b** R-T characteristics of VO$_2$ crossbar devices (active area: 750 nm × 750 nm × 60 nm) grown on a 1 μm SiO$_2$ substrate. The variability between the insulating and metallic-state resistances (ΔR$_{ins}$ = 2.36 × 10$^6$ Ω·μm and ΔR$_{met}$ = 9.22 × 10$^4$ Ω·μm, respectively) is marked. AFM imaging and Raman spectroscopy of the VO$_2$ sample RTA annealed with the conditions: Temp: 470 °C, O$_2$ pressure: 5 Pa, Annealing time: 10 min. **c** Schematic representations of a planar VO$_2$ device and a crossbar VO$_2$ device with preferential current paths in purple.

**VO$_2$ Grown By Rapid Thermal Annealing (RTA)**

The RTA tool offers O$_2$ pressure control at a level as low as the STA tool while providing much faster heating and cooling ramps. Table S3 (SI) provides a summary of the various tested recipes, emphasizing whether they meet the required specifications to obtain high-quality VO$_2$ films for our oscillators. These specifications include the presence of VO$_2$ stoichiometry confirmed by Raman spectroscopy, layer uniformity, and sample reproducibility.

The AFM measurements in Figure 2b indicate that the highly reproducible annealing recipe highlighted in Table S3 results in grains as small as those obtained with the slow annealer (see Figure 1a). Additionally, the surface is nearly as smooth as in samples subjected to flash annealing (Figure 2a) and the main Raman peaks of $VO_2$ (193 cm$^{-1}$ and 223 cm$^{-1}$) are detected using this recipe (Figure 2b)[39]. Figure 2b shows that the ratio of the intensities of the main Raman peaks to the weaker ones (389 cm$^{-1}$, 497 cm$^{-1}$, 612 cm$^{-1}$, is higher compared to those obtained with flash annealing (Figure 2a)[39]. This suggests a reduced surface roughness [39], which is confirmed by the measurements in Figure 2b. This annealing technique with the highlighted conditions in Table S3 was selected to realize the $VO_2$ oscillators.

Figure 2b shows the R-T characteristics of crossbar devices of identical dimensions (active area of 750 nm × 750 nm × 60 nm) fabricated on a chip with these conditions. Despite the ideal structural composition of the film and the smooth transitions observed in the hysteresis curves, which exhibit relatively sharp transitions with a narrow hysteresis window of less than 20 °C for a granular film[30,36], the devices show variability in their electrical properties. The measured resistivities range from $1.46 \times 10^6$ Ω·µm to $3.82 \times 10^6$ Ω·µm ($\Delta R_{ins} = 2.36 \times 10^6$ Ω·µm) in the insulator state, and from $1.88 \times 10^4$ Ω·µm to $1.11 \times 10^5$ Ω·µm ($\Delta R_{met} = 9.22 \times 10^4$ Ω·µm) in the metallic state. Given the unavoidable variability between $VO_2$ devices grown on $SiO_2$, we investigate, in the next section, the growth of $VO_2$ grains on different substrate stacks. The objective is to enhance material uniformity and device reproducibility by introducing an interlayer between the $VO_2$ and the $SiO_2$ substrate. By leveraging the diverse surface energies of these interlayers, we want to take advantage of the de-wetting angles created between the $VO_2$ grains and the underlying surface.

**$VO_2$ On Metal Oxides**

Figure 3a shows the normalized R-T characteristics of $VO_2$ layers in stacks embedding $HfO_2$, $Ti_3O_5$, $Al_2O_3$, or $WO_x$ interlayers (red, green, blue, and orange curve, respectively) annealed for 45 minutes at 520 °C, along with a reference sample of VO$_2$ on 1 μm of SiO$_2$ (black curve). These materials were selected due to their widespread availability in a semiconductor fab environment and their higher surface energies compared to silicon. The longer 45-minute annealing ensures the formation of de-wetted VO$_2$ grains on the interlayers. The presence of a thin 10 nm interlayer significantly alters the R-T characteristics compared to the reference sample (black curve). The VO$_2$ layer deposited on WO$_x$ exhibits a less steep transition compared to typical samples on SiO$_2$. On Al$_2$O$_3$, the VO$_2$ layer displays similar hysteretic characteristics to the reference SiO$_2$ sample, but with a slight shift towards higher temperatures, suggesting it requires higher voltages to trigger its phase transition. Similarly, the Ti$_3$O$_5$ interlayer results in an even higher transition temperature accompanied with a narrower hysteresis width. To achieve low-power operation, we discarded Al$_2$O$_3$ and Ti$_3$O$_5$ as suitable interlayers for the oscillators. Interestingly, the sample with VO$_2$ on HfO$_2$ annealed for 45 minutes shows an increase in the transition temperature of the IMT (80 °C) while keeping the MIT around 68 °C.

Figure 3b shows the Raman spectra corresponding to the samples with different interlayers. In samples with Al$_2$O$_3$ or Ti$_3$O$_5$ interlayer, a shift of either one of the two major peaks of VO$_2$ is observed, suggesting the presence of strain across the VO$_2$ layer. The characteristic VO$_2$ Raman peaks at 193 cm$^{-1}$ and 223 cm$^{-1}$ are sharp and more pronounced with the HfO$_2$ interlayer, indicating superior stabilization into the VO$_2$ oxidation state and more uniform crystallization compared to other interlayers[39]. This is an important finding showing that the incorporation of a 10 nm HfO$_2$ interlayer with high surface energy prevents the complete separation of VO$_2$ into individual nanocrystals during the annealing process, as observed in previous studies[6,44].

The results in Figure 3a-b indicate that long annealing times at high temperatures can be employed to tune the intrinsic properties of VO$_2$ layer when combined with a metal-oxide interlayer.

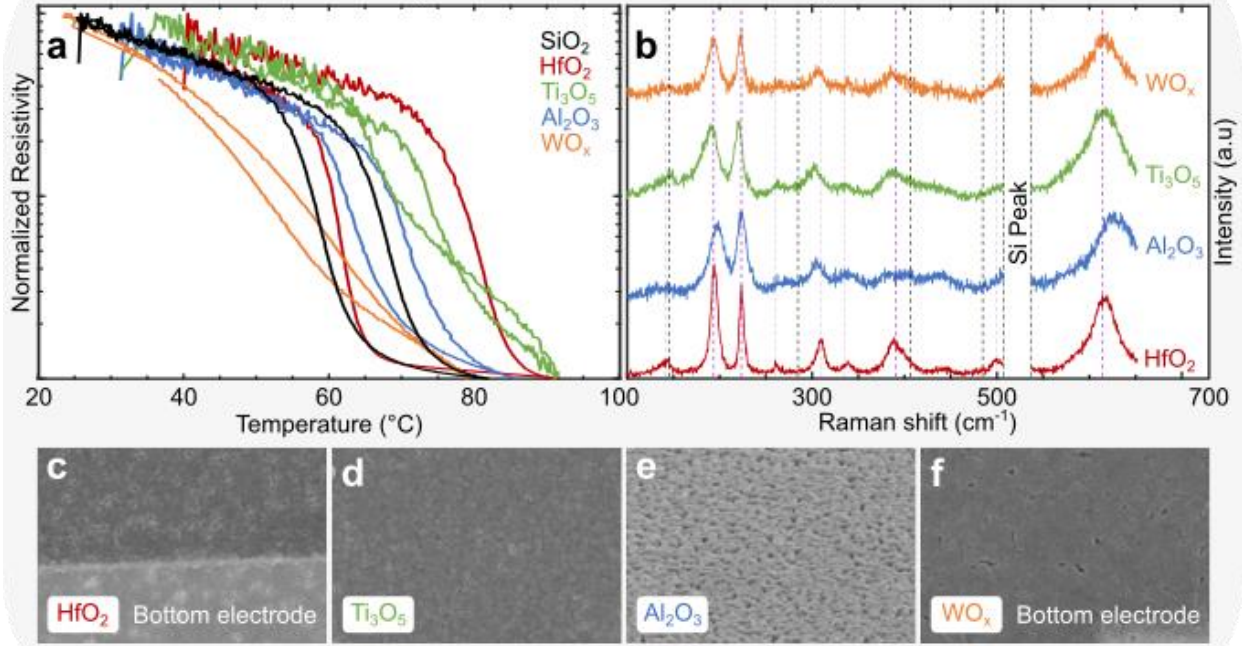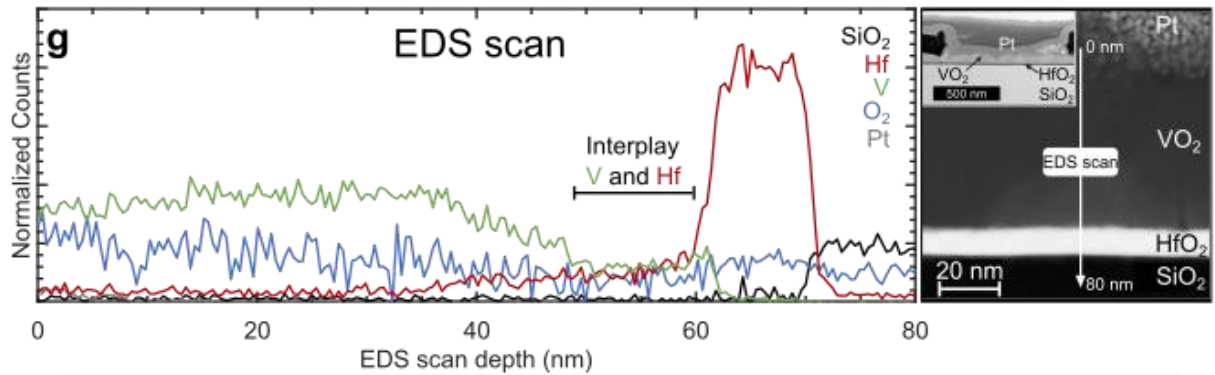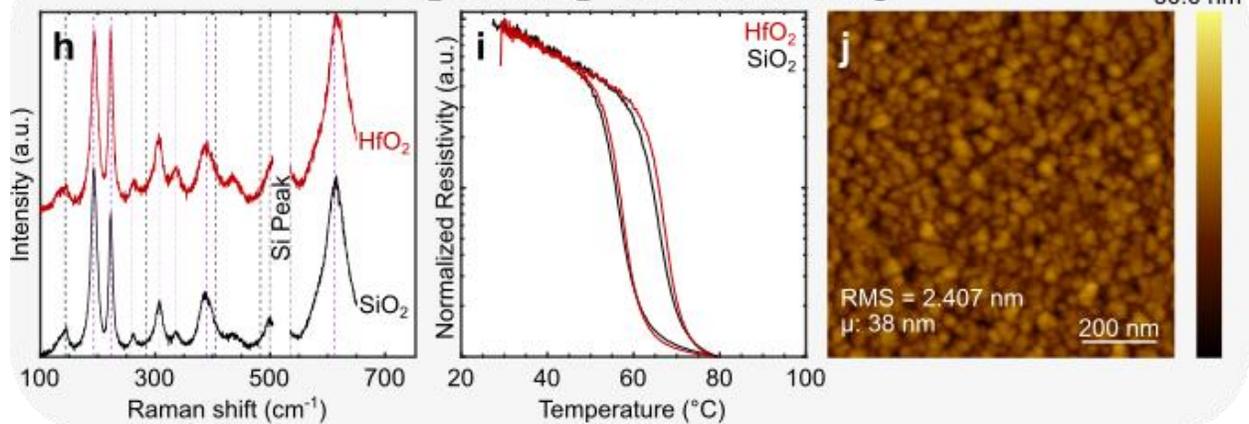

**Figure 3. VO$_2$ grown on interlayers. a** Normalized R-T and **b** Raman spectra for VO$_2$ samples with 1 μm thermal SiO$_2$ coated with 10nm interlayer of either HfO$_2$ (ALD), Al$_2$O$_3$ (ALD), Ti$_3$O$_5$ (Evaporated) or WO$_x$ (ALD), STA treated: 520 °C, 5 Pa O$_2$ pressure, 45 min long annealing. **c-f** SEM images of the samples. **g** EDS composition analysis at the interfaces of the sample in **c**, revealing that a thin layer of about 8 nm (located between 50 nm and 60 nm deep from the start of the EDS scan) shows some interplay between VO$_2$ and HfO$_2$. **h** Raman spectra, **i** normalized R-T, and **j** AFM measurement of VO$_2$ samples with 1 μm of thermal SiO$_2$ coated with either no interlayer (black) or a 10 nm interlayer (red) of HfO$_2$ (ALD), STA treated: 520 °C, 5 Pa O$_2$ pressure, 10 min long annealing.

To link the differences in R-T characteristics between samples to possible structural differences, we compared the SEM images of the VO$_2$ on HfO$_2$, Al$_2$O$_3$, Ti$_3$O$_5$, and WO$_x$ samples in Figure 3c-f. These images, along with the corresponding AFM measurements, reveal similar grain size and surface roughness in all samples with various interlayers, suggesting that the observed R-T differences in Figure 3a might be attributed to variations in the vanadium oxide layer at the interface with the underlying substrate.

To confirm this theory, we investigate the VO$_2$ to HfO$_2$ interface for the sample annealed for 45 minutes at 520 °C, as it shows the largest variation in transition temperature compared to the reference sample on SiO$_2$ substrate (Figure 3a, red and black lines, respectively). In Figure 3g, the transmission electron microscope (TEM) micrograph and energy dispersive spectroscopy (EDS) scans measured on the sample with the HfO$_2$ interlayer highlight the formation of a thin layer at the VO$_2$/HfO$_2$ interface of roughly 8 nm, where both Hafnium and Vanadium are present. This interplay is a possible cause for the increased IMT (red curve in Figure 3a), bringing more evidence that metal-oxides such as HfO$_2$ can be combined with long annealing times at high temperatures to engineer the crystalline phase-transition of $VO_2$. While this constitutes an interesting finding, the primary focus remains on minimizing variability among insulating- and metallic-state resistances, as illustrated in Figure 1b and Figure 2.

Based on the intensity and sharpness of the Raman peaks in Figure 3b, $HfO_2$ emerges as the best interlayer to crystallize high-quality $VO_2$ grains. Continuing with this interlayer, we now use shorter annealing times to avoid intermixing caused by long annealing at the interface and maintain the IMT close to 68 °C.

Figure 3i shows the R-T characteristics of $VO_2$ annealed for 10 minutes on a $HfO_2$ interlayer and of a reference sample with no interlayer. With this short annealing time, the presence of $HfO_2$ does not influence the hysteretic width or transition temperatures of $VO_2$. More importantly, the variability in resistivities observed on the samples grown directly on $SiO_2$ (reported in Figure 1b and Figure 2) is greatly reduced across the $VO_2$ layer when an $HfO_2$ interlayer is present (see Figure 7a-c). This leads to the reproducible characteristics presented in the following section. The AFM measurements in Figure 3j show that the average grain size (38 nm) remains small and the surface, smooth (2.4 nm) in the presence of $HfO_2$; a finding also reported in Zong *et al*.[45]

In summary, $HfO_2$ is the material with the most interesting effect on the formation of $VO_2$. It preserves its fundamental switching properties in the case of short annealing times (10 minutes) while achieving high crystalline quality across the entire layer, even without using the optimized RTA annealing technique. In the next section, we perform an XRR and TEM study to explore the reasons why incorporating $HfO_2$, along with a very thin $SiO_2$ layer, reduces variability, and thus constitutes the way forward to successfully attain the level of uniformity required to couple $VO_2$-based oscillators.

**Substrate Thickness Dependence Study**

Figure 4a-b shows the XRR patterns of samples *D* and *E* characterized by a vanadium oxide layer deposited on 1 μm SiO$_2$, before and after annealing respectively, and Figure 4c, of sample *F* using a HfO$_2$ interlayer after annealing. The XRR measurements are analyzed by fitting a simulated curve, based on a multilayer model, to the measured data[46]. The structural properties such as density, layer uniformity, and thickness are then compared and linked to the electrical performance of the various samples. Additional information regarding the model's efficacy and sensitivity can be found in the SI. The fits are in good agreement with the experimental datapoints and the outcome of the analysis is reported for all three samples (*D*, *E*, and *F*) in Table 1. For the as-deposited sample *D*, vanadium oxide does not grow as a single uniform layer. A VO$_x$ layer with a density of 3.9 g cm$^{-3}$, consistent with amorphous vanadium oxide densities, is followed by a second layer with an even lower density of 3.2 g cm$^{-3}$ directly in contact with the SiO$_2$ [47]. Following the annealing of sample *D*, the XRR analysis of sample *E* shows that, while it crystallized into grains (as evidenced by the increase in density to 4.6 g cm$^{-3}$ in the top layer), the unwanted spurious layer persists and grows thicker, with its density slightly increasing from 3.2 g cm$^{-3}$ (prior to annealing) to 3.9 g cm$^{-3}$ [47]. This can be attributed to a non-uniform clustering of the material at the VO$_2$/SiO$_2$ interface to form grains of VO$_2$. The lower density layer cannot be visualized sharply when imaging the samples with the local TEM technique shown in Figure 5a-c, but it is reflected in several Fast Fourier transform (FFT) plots captured at the interface between the substrate and the VO$_2$ grains (Figure 5e and f). Despite the grains being mostly found in a single orientation crystallization as in Figure 5d and g, amorphous material can still be found at the grain boundaries. The XRR measurement, which measures a wider portion of the sample, suggests that the amorphous material at the grain boundaries extends on average across the whole VO$_2$/SiO$_2$

interface. These findings consistently extend to similar samples with intermediate SiO$_2$ thicknesses down to 50 nm, as discussed in SI. This could also account for the observed differences in resistivity ranges across our samples (Figure 1b and Figure 2), as the presence of this spurious layer may introduce varying series resistance levels from one sample to another, causing a vertical shift in the R-T characteristics. The introduction of an HfO$_2$ interlayer between VO$_2$ and SiO$_2$, as in sample *F*, results in more uniform growth of VO$_2$ grains across the entire film thickness, as shown by XRR analysis in Table 1. This observation is consistent with previous studies[45,48] suggesting that HfO$_2$ can regulate VO$_2$ crystallinity by promoting the nucleation of grains and reducing boundary defects. Interestingly, in VO$_2$ films grown on ultrathin (≤10 nm) SiO$_2$ layers on Si substrates, no spurious interfacial layers are detected and the structural properties of the crystallized VO$_2$ layer compare well with the sample presenting the HfO$_2$ interlayer. The results are summarized in SI.

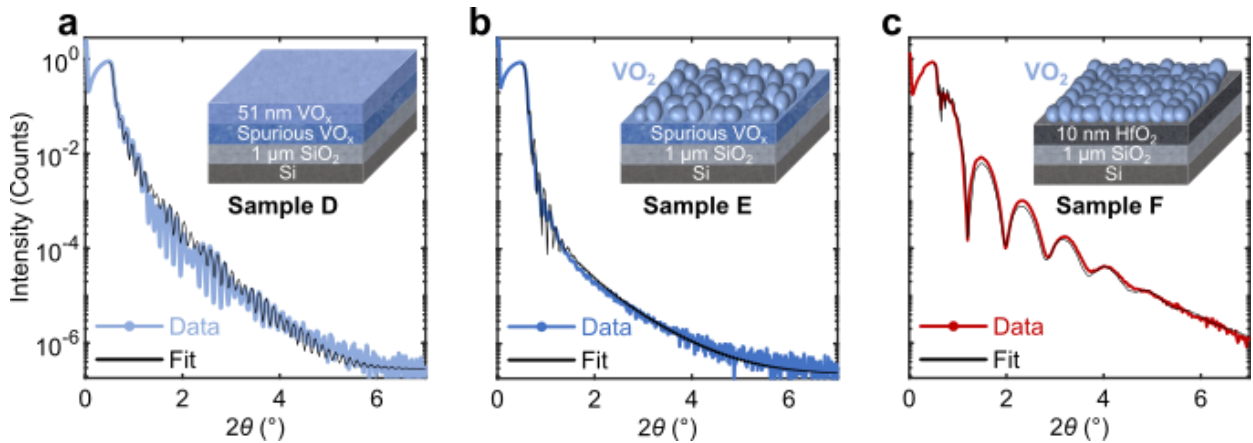

**Figure 4. XRR measurements.** The experimental data (blue line) correspond to a vanadium oxide layer deposited on 1 μm SiO$_2$ a before annealing (Sample *D*) and b after annealing (Sample *E*). c XRR patterns of a VO$_2$ layer annealed on a 1 μm SiO$_2$ substrate with a 10 nm thick HfO$_2$ interlayer (Sample *F*). Simulation curves (black line) obtained from the multilayer model stacks shown in each figure are used to fit the experimental curves.

**Table 1. XRR fit analysis for samples *D*, *E*, and *F*.**

| Sample Stack | Thickness (nm) | Roughness (nm) | Density (g cm$^{-3}$) |
|---|---|---|---|
| *Sample D* | | | |
| Unannealed VO$_x$ | 51.4 | 0.5 | 3.9 |
| Spurious VO$_x$ layer | 9.6 | 0.6 | 3.2 |
| SiO$_2$ | 1000 | 0.5 | 2.3 |
| *Sample E* | | | |
| Annealed VO$_2$ | 43.2 | 3.0 | 4.3 |
| Spurious VO$_x$ layer | 19.0 | 2.5 | 3.6 |
| SiO$_2$ | 1000 | 0.3 | 2.3 |
| *Sample F* | | | |
| Annealed VO$_2$ | 51.7 | 2.7 | 4.1 |
| HfO$_2$ interlayer | 9.9 | 0.6 | 10.5 |
| SiO$_2$ | 1000 | 0.3 | 2.3 |

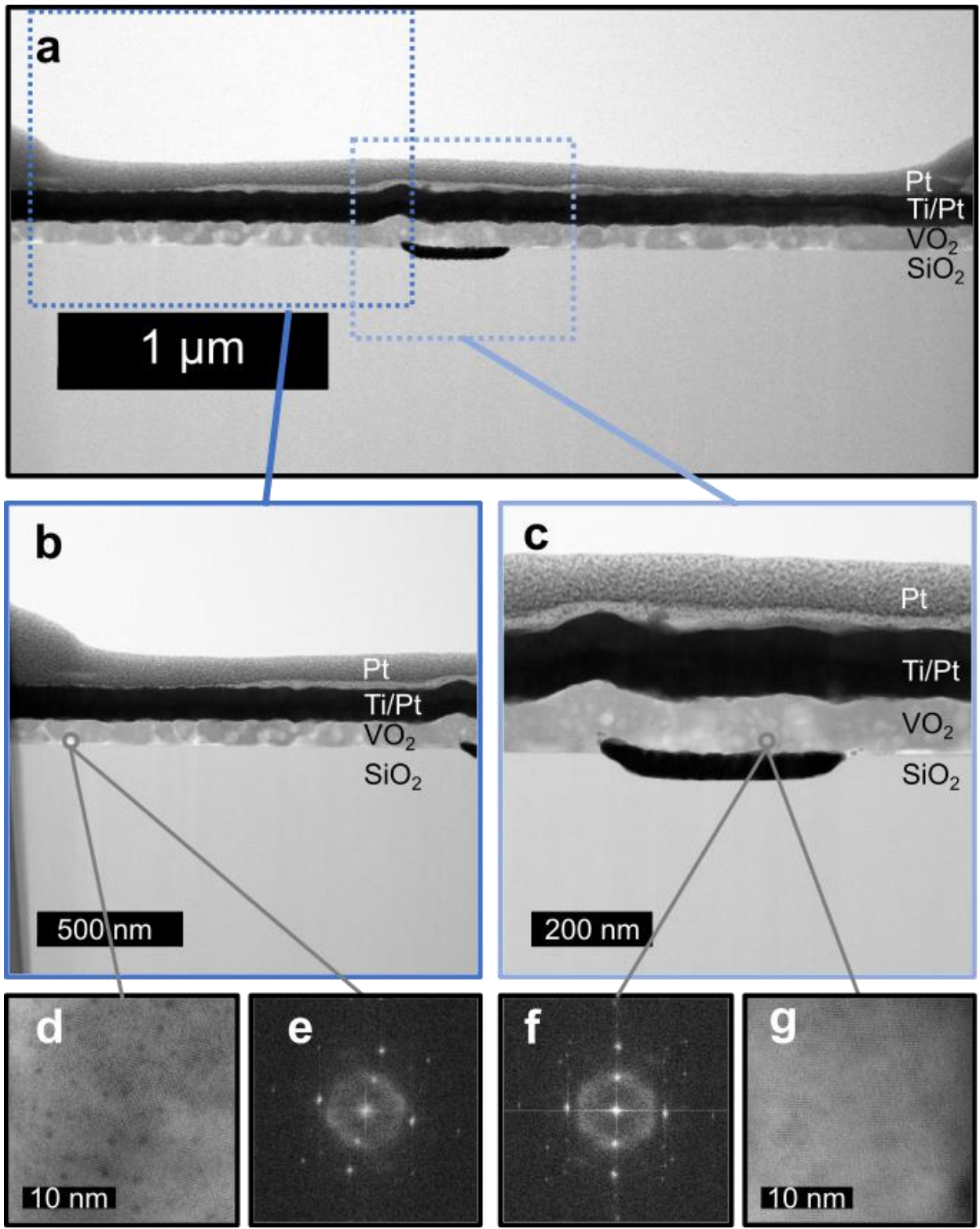

**Figure 5. TEM analysis of VO$_2$ grown on SiO$_2$. a** TEM image of a crossbar device with VO$_2$ grains. **b** Grains grow directly on the 1 μm SiO$_2$ substrate or **c** the bottom electrode. Single-layer

VO$_2$ grains show interfacial crystallization **d,g** at the VO$_2$/SiO$_2$ interface in only one orientation in the **e,f** FFT plots. The cloud-shaped rings around the origins of the FFT plots reveal the presence of leftover amorphous vanadium-oxide material.

The absence of this spurious interfacial layer leads to a low variability between devices, measured in Figure 6c in comparison with those in Figure 6b. In the Raman spectra of Figure 6a, characteristic VO$_2$ peaks are observed on both thin (≤10 nm) or thick (1 μm) SiO$_2$ substrates. However, only devices fabricated on thin SiO$_2$ or with a HfO$_2$ interlayer, which are both characterized by the absence of the additional amorphous spurious layers (Figure S1 in SI and Figure 4c, respectively), demonstrate reproducible and uniform electrical properties (see Figure 6c and Figure 7a-c, respectively). This suggests that by biasing the VO$_2$ grains in such devices, despite the presence of Ti and Pt in crossbar devices, a preferential current path will be established from the top to the bottom electrodes, avoiding any interference from the spurious layer, effectively removing variability between devices observed in Figure 1c, Figure 2, and Figure 6b.

In the crossbar configuration, favored for precisely defining the current path in the device, a thin (≤10 nm) SiO$_2$ is not recommended, as it may result in the electrodes coming into contact with the semiconducting Si substrate after patterning. This can be avoided by using HfO$_2$ as an etch-stop layer. Therefore, the role of the HfO$_2$ interlayer is threefold: **1.** It acts as a barrier to avert the formation of this undesired spurious layer, responsible for creating variability among VO$_2$-based crossbar devices, **2.** It preserves VO$_2$'s structural and phase-transition properties, while **3.** acting as an etch-stop layer between the semiconducting Si substrate and the VO$_2$ grains for further processing and ohmic contacts.

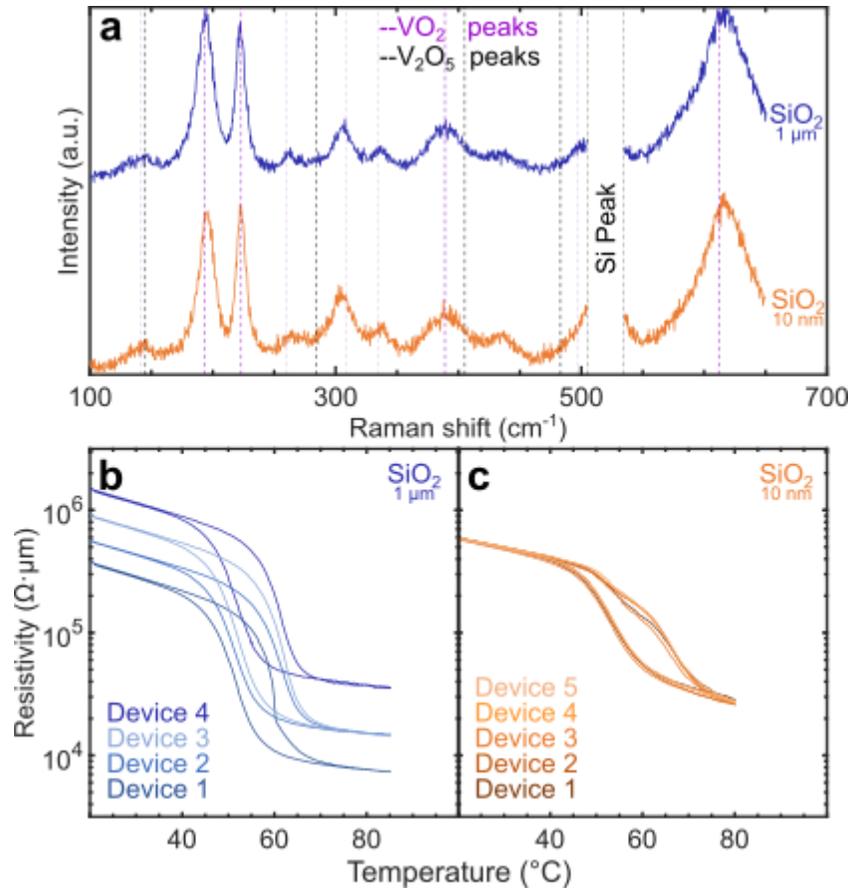

**Figure 6. Influence of the SiO$_2$ thickness on VO$_2$. a** Raman spectra for STA treated VO$_2$ samples on SiO$_2$ substrate layer grown thermally (1 μm) or by PECVD (10 nm). Differentiating VO$_2$ grains crystallinity and quality cannot be assessed through Raman analysis as the spectra are indistinguishable. R-T characteristics of VO$_2$ crossbar devices grown on **b** a 1 μm thermal SiO$_2$ substrate and **c** a 10 nm SiO$_2$ (PECVD) substrate. Both the insulator- and the metallic-state resistivities show significant device-to-device variability on thick SiO$_2$ substrate **b** compared to the ones on a thin SiO$_2$ layer **c**.

### VO$_2$ Devices for Relaxation Oscillators

The material development leads to these findings: the highest quality of VO$_2$ layer is obtained with a thin SiO$_2$ (≤10 nm)-HfO$_2$ (10 nm)-VO$_2$ stack annealed under the conditions described in Table S4. The reproducibility of these devices makes them the best choice to build relaxation oscillators. The crossbar oscillators are designed with 50 nm thick Ti/Pt top and bottom electrodes. The active area has dimensions between 100 nm × 100 nm × 60 nm and 300 nm × 300 nm × 60 nm.

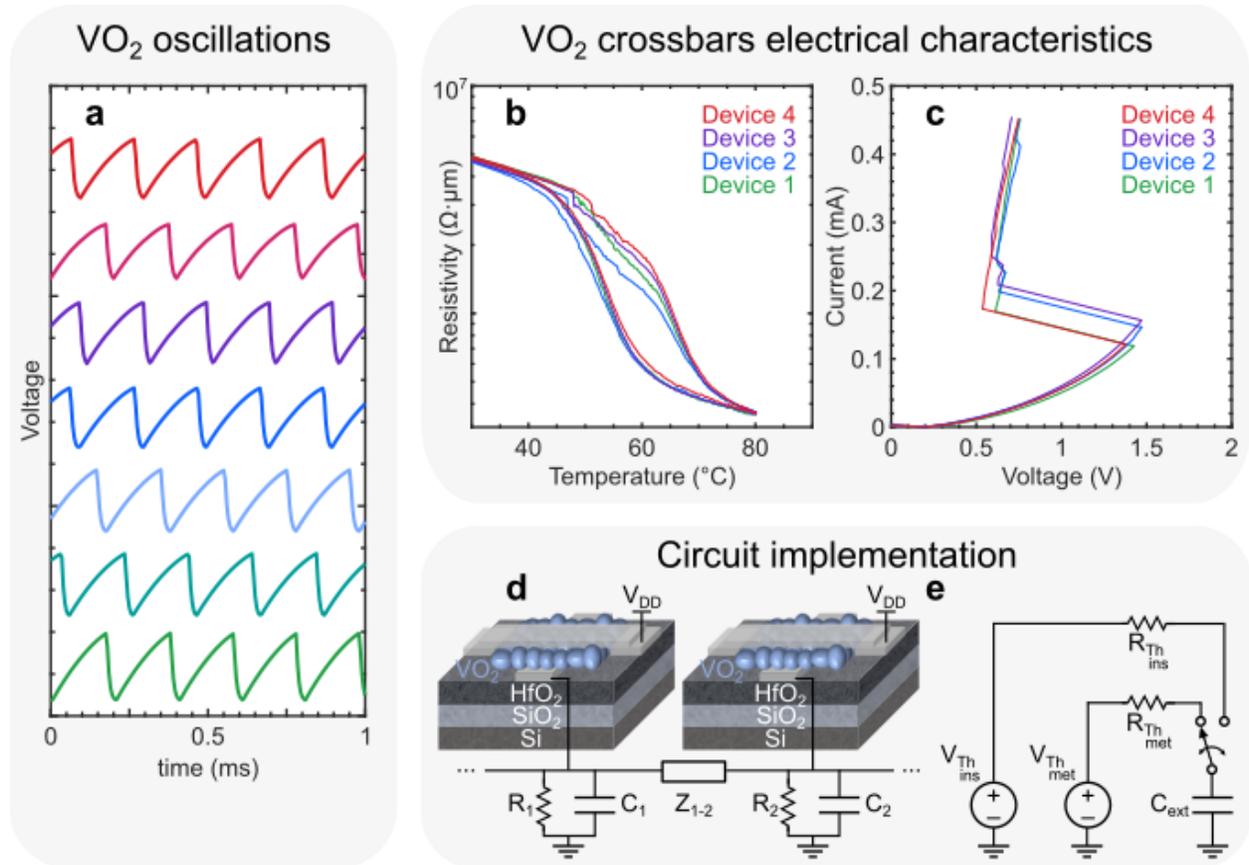

**Figure 7. VO₂ oscillators. a** Oscillatory response of 7 VO$_2$-based crossbar devices biased with a DC voltage. The traces are stacked to facilitate visibility. $V_{DD}$ = 5 V. $R_s$ ≈ 40 kΩ. $C_{ext}$ = 10 nF. **b** R-T characteristics of VO$_2$ crossbar devices grown on a thin SiO$_2$ (≤10 nm)-HfO$_2$ (10 nm) stack, RTA annealed: 470 °C, 5 Pa O$_2$ pressure, 10 min long annealing. **c** Current-Voltage (I-V) characteristics at room temperature of 4 VO$_2$-based crossbar devices biased with a controlled ramping voltage from. $V_{DD}$= 0 V to 10 V. $R_s$≈40 kΩ. The samples have 4 nm of PECVD SiO$_2$, 10 nm of HfO$_2$ (ALD), and 60 nm of annealed VO$_2$. (active area: 300 nm × 300 nm × 60 nm). **d** Schematic of VO$_2$-based oscillators built with a crossbar device, an external resistor and a capacitor to set and adjust the oscillating frequency. The coupling units $Z_{i-j}$ connect each oscillator's output to create an ONN, whose phase relationships define a state of encoded information. **e** Thévenin equivalent circuit of one $VO_2$-based oscillator seen from the output terminal of an external load capacitor.

Figure 7b shows the R-T characteristics of four devices fabricated on a thin $SiO_2$ (≤10 nm)-$HfO_2$ (10 nm) stack annealed under the conditions highlighted in Table S4. Figure 7c shows the corresponding I-V characteristics where $V_{DD}$ was swept from 0 V to 10 V, with a series resistor of 40 kΩ connected to the bottom electrode, and no external load capacitor. When the voltage across each device reaches their respective insulator-to-metal transition point, i.e. when the material's temperature has reached 68 °C, the current and the conductivity increase abruptly. We can conclude from our measurements in Figure 7a-c and in SI that the insulator- and metallic-state resistivities are nearly identical for all devices, with transition voltages $V_{IMT}$ varying by less than 7%. These results confirm the mitigated variability between the oscillators achieved with our material and stack optimization. The low transition from the insulator- to the metallic-state of about 1.5 orders of magnitude aligns with expectations for granular devices produced on a CMOS-compatible platform[30]. In such devices, the conductance of the metallic-state is constrained by grain boundaries, even when individual grains exhibit low resistance[27]. Consequently, the operational window, particularly the selection range for the $V_{DD}$ and $R_s$ parameters[49] (see Figure 7d), is narrower compared to devices made from single-crystal or epitaxially grown $VO_2$[36].

Figure 7d shows the circuit connections made to realize $VO_2$-based oscillators, including a Thévenin equivalent circuit in Figure 7e to model single device operation. The phase-change properties of $VO_2$ influence the equivalent circuit seen at the outputs. This behavior is represented in Figure 7e by a switch triggered by either the insulator-to-metal transition (IMT) or the metal-to-insulator transition (MIT). The Thévenin equivalent circuit also describes the charging and discharging phases at the output, allowing for the adjustment of the oscillation frequency by modifying the *RC*-network's equivalent time constant through the external series resistance. During the charging and the discharging phases of the capacitor, the VO$_2$ crossbar device is in the metallic state ($R_{VO_2} = R_{met}$) or the insulator state ($R_{VO_2} = R_{ins}$), respectively. The output voltage, measured at the terminal of the external capacitor, is expressed in two phases:

$$V_{out} = \begin{cases} V_{out_{charge}} = V_{TH_{met}} - [V_{TH_{met}} - (V_{DD} - V_{IMT})]e^{-t_{charge}/R_{TH_{met}}C_{ext}} \\ V_{out_{discharge}} = V_{TH_{ins}} + [(V_{DD} - V_{MIT}) - V_{TH_{ins}}]e^{-t_{discharge}/R_{TH_{ins}}C_{ext}} \end{cases} \quad (1)$$

With

$$V_{TH} = V_{DD}\frac{R_s}{R_s + R_{VO_2}} = \begin{cases} V_{TH_{met}} = V_{DD}\frac{R_s}{R_s + R_{met}} \\ V_{TH_{ins}} = V_{DD}\frac{R_s}{R_s + R_{ins}} \end{cases} \quad (2)$$

$$R_{TH} = \left(\frac{1}{R_s} + \frac{1}{R_{VO_2}}\right)^{-1} = \begin{cases} R_{TH_{met}} = \left(\frac{1}{R_s} + \frac{1}{R_{met}}\right)^{-1} \\ R_{TH_{ins}} = \left(\frac{1}{R_s} + \frac{1}{R_{ins}}\right)^{-1} \end{cases} \quad (3)$$

And ensuring that

$$V_{TH_{met}} > V_{DD} - V_{IMT} \quad (4)$$
$$V_{TH_{ins}} < V_{DD} - V_{MIT} \quad (5)$$

By fixing the external capacitor to C$_{ext}$ = 10 nF and imposing a desired oscillating frequency *f* of:

$$f = \left(t_{charge} + t_{discharge}\right)^{-1} = 5 \text{ kHz} \tag{6}$$

We can solve numerically Equation 1 and Equation 6 to find the required value of the external series resistance to operate at the set frequency. This method was applied to seven devices contacted simultaneously. The individual $VO_2$-based oscillators can be connected at their outputs by coupling units to mimic symmetrical synaptic weights – represented by impedances $Z_{i-j}$ in Figure 7d – to create an oscillating neural network.

Figure 7a shows the oscillations of seven devices oscillating at 5 kHz, with no coupling units (see Figure S7 in SI for Fast Fourier Transforms). The $VO_2$-based oscillators have similar oscillating voltage amplitudes and frequencies, indicating once more the low variability obtained with our annealing technique and material stack optimization investigations[50]. In Figure 7a, the repeatability in the R-T and I-V measurements shown in Figure 7b-c is demonstrated across several cycles of seven samples, revealing their similar oscillatory behaviors. The oscillation amplitude varies by less than 3.5% over 250 cycles within the same device, while the transition voltages show variations of up to 10% from one device to another. This degree of variability is at the boundaries of tolerance required for device coupling and enabling ONN-based computing[50]. In order to effectively realize an ONN, we opt for crossbars with device-to-device variability below 5%, as they tend to synchronize more easily. Designing such a network was not feasible with the characteristics of the devices shown in Figure 1b. In the case of high variability among the oscillators, they would either ignore each other in the case of weak coupling and not lock in frequency, or exchange too much current when coupled strongly, often leading to oscillation failure or device breakdown. An example is shown in SI.

**DISCUSSION**

Here, we discuss the role of interlayers and grain boundaries on the electrical performance of the oscillators. The TEM micrographs and EDS scan shown in Figure 3g reveal a possible interdiffusion mechanism occurring between the interlayer and the $VO_2$ at temperatures above 400 °C. Among the tested interlayers, $WO_x$ is the only material higher than vanadium oxides on an Ellingham diagram at the annealing temperatures[51,52]. $VO_2$ thus acts as a scavenging material, absorbing oxygen both from the $WO_x$ interlayer and the chamber. This interaction with a dopant or a substrate material can promote the formation of undesirable $VO_2$ crystal states (e.g. $VO_2(B)$) or the stabilization of uncontrolled oxidation states, leading to the degradation or suppression of the film's transition temperature (see Figure 3a). Similar results were also observed in Guo *et al*.[25] Interestingly, the same failure mechanism observed with the $WO_x$ interlayer is not seen for samples with $HfO_2$ or $Al_2O_3$ interlayers under long annealing (see Figure 3a). According to other studies[28,29,34,37], treating the film with such conditions should have stabilized the films in the $V_2O_5$ stoichiometry. Instead, we observed that doping $VO_2$ with $HfO_2$ or $Al_2O_3$ interlayers combined with long annealing produces phase-change materials with higher transitions temperatures. Our results presented in Figure 3a-f indicate that a broad range of substrates with various thicknesses can be harnessed to engineer the IMT in $VO_2$ films[53]. This opens new possibilities for engineering and tuning $VO_2$'s transition temperature, particularly for applications requiring a higher thermal budget. However, considering the specific application targeted in this study, the incorporation of interlayers other than $HfO_2$ or the direct growth of $VO_2$ grains on thick $SiO_2$ results in devices with unpredictable oscillation patterns. Due to challenges related to reproducibility, stability, and lack of configurability through external parameters ($R_s$ and $C_{ext}$), extensive measurement on these devices becomes impractical, and coupling is impossible. See SI.

We now discuss the role of oxygen in annealing stable $VO_2$. We observed a significant reduction in sample-to-sample variability by reducing the $SiO_2$ thickness (Figure 6c), which points to a potential oxygen diffusion mechanism between the substrate and the vanadium oxide layer during the early stages of the annealing step[54]. This diffusion process could contribute to the stabilization into the $VO_2$ oxidation state and effectively prevent excess oxygen from forming higher oxidation states such as $V_2O_5$ crystals. In fact, our measurements in Figure 6c and Figure 7a-c suggest that when the vanadium oxide film can lose oxygen through a thin $SiO_2$ layer and the $HfO_2$ interlayer into the Si substrate below[55], our optimized RTA annealing consistently produces samples of high quality, a finding also reported in Prasadam *et al.*[29] In the case of a thick $SiO_2$ substrate with no interlayer, the oxygen, unable to diffuse into the Si substrate, may remain at the interface between the $VO_2$ grains and the substrate. This is a meaningful discovery, as we believe that the oxygen excess at the grain boundaries near the $VO_2/SiO_2$ interface, detected by the main $V_2O_5$ Raman peak (145 cm$^{-1}$) in Figure 3h and Figure 6a, is responsible for creating the measured spurious layer (see Figure 4 and Table 1) that led to variability among samples (Figure 1b, Figure 2, and Figure 6b).

Here, we explain this variability by analyzing the impact of an electric field at the nanoscale level. From a device standpoint, applying a forward or backward bias respectively generates or annihilates oxygen vacancies that trigger the heterogeneous nucleation of the phase transition, thus influencing the resistance state of the $VO_2$ film[23,56]. In the case of devices on thick $SiO_2$ substrate, which exhibit more defects and dislocations, the non-uniform diffusion of oxygen during annealing and/or the presence of a disordered layer with higher oxygen density at the interface (Table 1) lead to uneven conductance at the grain boundaries. Consequently, biasing $VO_2$ on a thick $SiO_2$ substrate results in non-uniform oxygen vacancies at the grain boundaries, which translates into the varying insulator- and metallic-state resistivities (Figure 1b, Figure 2, and Figure 6b). This effect is exacerbated with our dense polycrystalline layer, as the number of grain boundaries increases inversely with the size of the $VO_2$ grains. This further motivates our choice of a thin $SiO_2$ layer (≤10 nm)-$HfO_2$ (10 nm) stack that effectively reduces variability in the oscillators by avoiding the formation of a spurious layer during ALD deposition (Figure 4a). This allows the current path to follow the grain boundaries of the monolayer $VO_2$ between the top and bottom electrodes without crossing the spurious layer. As a result, the $VO_2$ grown on this optimized stack configuration (Figure 4c) and using our best annealing technique (Table S3) leads to a more uniform generation of oxygen vacancies at the grain boundaries during electrical operation.

## CONCLUSIONS

The realization of a large-scale network comprising several $VO_2$-based oscillators has been hindered by the challenging stabilization of $VO_2$ oxidation state, typically introducing granularities and rough morphology. This has led to degraded electrical performance and significant variability between devices, generally limiting the coupling to only two devices. To achieve the required device reliability for a neuro-inspired circuit, we investigated the role of annealing parameters using three different methods post-ALD deposition of the vanadium oxide film. Our goal was to obtain a high-quality $VO_2$ layer with low surface roughness, densely positioned small grains, and highly reproducible current-voltage (I-V) and hysteretic R-T characteristics.

To achieve this level of quality, we employed a rapid thermal annealing technique capable of delivering a quick and uniform heat distribution across the entire vanadium oxide layer. Our findings revealed that device variability was attributed to the formation of an amorphous spurious vanadium oxide layer at the $VO_x/SiO_2$ interface when $SiO_2$ was thick (>10 nm). This layer significantly affected the generation of oxygen vacancies during device operation, leading to an uncontrolled current path within the cross-section area of our crossbar $VO_2$-based oscillators.

Furthermore, we studied the role of different metal-oxide interlayers placed between the $Si/SiO_2$ substrate and the $VO_2$. By engineering a stack that included a $HfO_2$ interlayer, we obtained a sharper interface, resulting in mitigated device variability. By combining our highly reproducible annealing treatment with our optimized epitaxial stack $SiO_2$ (≤10 nm) – $HfO_2$ (10 nm) underneath the vanadium-oxide layer, we achieved excellent results, with up to 7 VO2-based oscillators simultaneously contacted on a CMOS medium. These oscillators operated at the exact same frequency, with oscillation amplitudes on the order of 1.7 V. This level of uniformity and ideal electrical performance meets the requirements for successful device coupling and the realization of an oscillating neural network. Our network of $VO_2$-based oscillators presents an attractive and scalable computing unit for hardware accelerators, offering new computational paradigms for AI applications in optimization problems and pattern recognition, thanks to its high-performance switching properties and CMOS compatibility[5,7,11].

## DATA AVAILABILITY

Datasets generated during the current study are available from the corresponding authors on request. More XRR measurements, edge effects caused by flash annealing, effect of the $SiO_2$ growth method on the growth of granular $VO_2$, typical problems encountered when coupling $VO_2$ oscillators presenting high device-to-device variability, AFM measurement values, FFT on the devices' oscillations, and RTA conditions tested are available in SI; Figures S1-S8 and Tables S1-S3.

## AUTHOR INFORMATION

**Corresponding Author**

Olivier Maher – IBM Research Zurich; orcid.org/0009-0005-7846-1493Email: OGM@zurich.ibm.com

Siegfried Karg – IBM Research Zurich; Email: SFK@zurich.ibm.com

**Author Contributions**

O. Maher, R. Bernini, N. Harnack, M. Sousa, and V. Bragaglia collected the experimental datasets presented in the manuscripts. The manuscript was written through contributions of all authors. B. Gotsmann and S. Karg supervised and directed the project. All authors have given approval to the final version of the manuscript.

**Funding Sources**

This project has received funding from the EU's Horizon program under projects No. 871501 (NeurONN), 101092096 (PHASTRAC), and No. 861153 (MANIC).

**ACKNOWLEDGMENT**

This project has received funding from the EU's Horizon program under projects No. 871501 (NeurONN), 101092096 (PHASTRAC), and No. 861153 (MANIC).

The authors thank the Cleanroom Operations Team of the Binnig and Rohrer Nanotechnology Center (BRNC) for their help and support.

**COMPETING INTERESTS**

The authors declare no competing interest.

# Highly Reproducible and CMOS-compatible $VO_2$-based Oscillators for Brain-inspired Computing

*Olivier Maher\*, Roy Bernini, Nele Harnack, Bernd Gotsmann, Marilyne Sousa, Valeria Bragaglia, and Siegfried Karg\**

IBM Research Zurich, Säumerstrasse 4, 8803 Rüschlikon, Zürich, Switzerland

SUPPORTING INFORMATION

**XRR reflectivity measurements**

The XRR curves of Figure S1 are acquired with a Bruker D8 discover diffractometer equipped with a rotating anode generator and analyzed by fitting a simulated curve, based on a multilayer model, to the measured data[1]. XRR profile of Sample K with nominal 50 nm of amorphous $VO_2$ deposited on ultrathin 2 nm Si substrate and post annealed to obtain a polycrystalline $VO_2$ layer. The XRR analysis demonstrate that the sample has sharp interfaces and no spurious layer between $VO_2$, and Si layers is formed as reported in Table S1. Sample L, instead, has nominal 50 nm of amorphous $VO_2$ films grown on 50 nm $SiO_2$ layer on Si substrate. The XRR analysis reveals that a spurious interfacial layer of ~9 nm with low density of ~3.5g cm$^{-3}$ is detected between the top $VO_2$ layer and the $SiO_2$. Sample M is similar to Sample K but with 10 nm $HfO_2$ interlayer. Also in this case, no spurious interfacial layers are obtained between the top $VO_2$ and underneath $HfO_2$ layer. All results are summarized in Table S1.

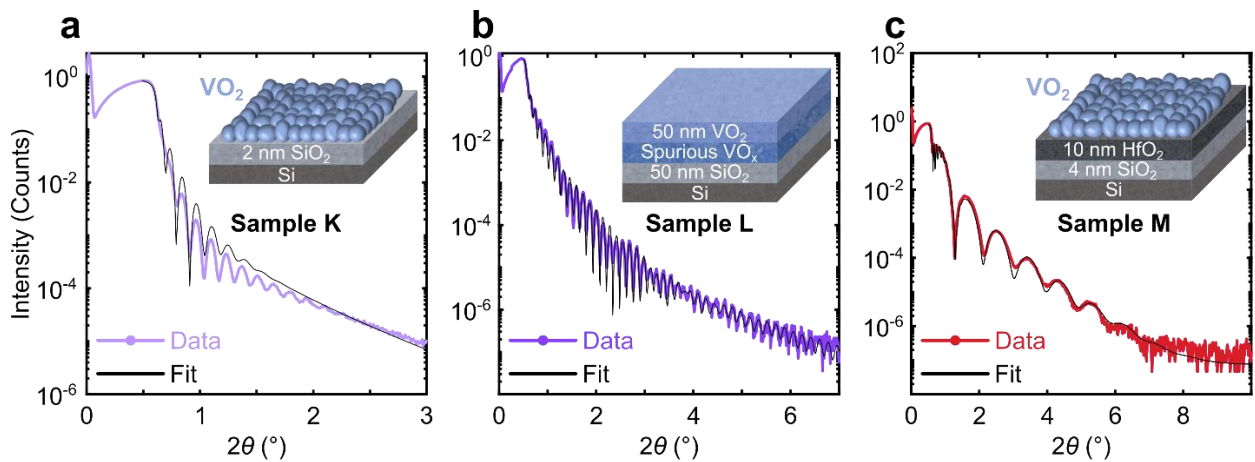

**Figure S1. XRR patterns of vanadium oxide samples.** The various stacks are characterized by (**a**) a vanadium oxide layer deposited on 2 nm SiO$_2$ after annealing (Sample K); (**b**) vanadium oxide deposited on 50 nm SiO$_2$ before annealing (Sample L) and (**c**) vanadium oxide deposited on HfO$_2$ layer after annealing (Stack M). Fits are in black while the experimental data are in colored line according to the legend of each figure.

**Table S1. XRR fit analysis for stacks K, L, and M.**

| Sample | Sample Stack | Thickness (nm) | Roughness (nm) | Density (g cm$^{-3}$) |
|---|---|---|---|---|
| **Sample K** | Annealed VO$_2$ | 52.6 | 2.9 | 4.6 |
| | SiO$_2$ | 2.0 | 0.5 | 2.3 |
| **Sample L** | Unannealed VO$_x$ | 52.8 | 0.9 | 3.8 |
| | Spurious VO$_x$ layer | 9.0 | 0.5 | 3.5 |
| | SiO$_2$ | 48.9 | 0.2 | 2.2 |
| **Sample M** | Annealed VO$_2$ | 62.4 | 2.4 | 4.3 |
| | HfO$_2$ interlayer | 9.1 | 0.7 | 11.5 |
| | SiO$_2$ | 3.9 | 0.4 | 2.3 |

**Flash annealing edge effect**

Figure S2 shows the gradual response of the film to the flash annealer across areas going from the center (α) to the edge (δ) of a sample annealed with a flash power of 90 J cm$^{-2}$, an oxygen partial pressure of 20 Pa, and pre-heated at 245 °C. The non-uniform heat distribution generated by the tool lamp introduced edge effects that translated into visible concentric rings on the surface, whose boundaries defined areas where grain size, surface roughness, and vanadium oxidation states changed drastically (see Figure S2c). Close to the center (α), small dense VO$_2$ grains were detected, progressively making way to larger V$_2$O$_5$ microsized grains towards the edges (δ), measured by Raman spectroscopy. In Figure S2a, the grain size and oxidation state variations across sample areas (α, β, and δ) directly relate to the varying R-T response, affecting transition temperatures, resistivities, and hysteresis widths.

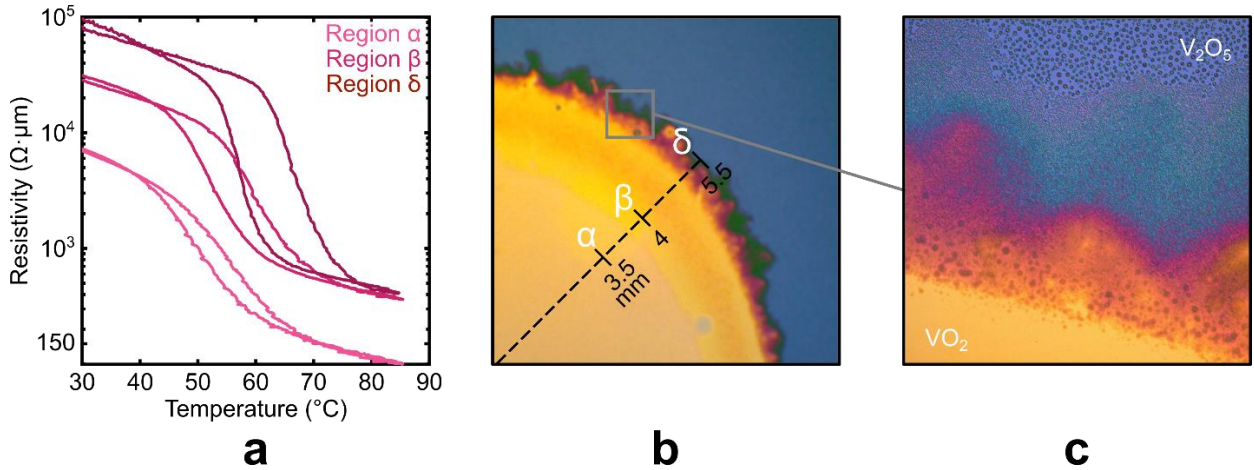

**Figure S2. Flash annealing edge effect.** (**a**) 4-probe R-T characteristics measured in different regions (α, β, and δ) of a sample FLA-annealed with the conditions: Pre-flash temp: 245 °C, $O_2$ pressure: 20 Pa, Annealing time: 20 ms, Flash power: 90 J cm$^{-2}$. (**b**) Microscopic view of the film showing the varying response to the flash. The progressively larger vanadium-oxide grains are visible in (**c**).

**VO₂ Growth On Silicon Dioxide (SiO₂)**

Table S2 summarizes the study results about the SiO$_2$ substrate influence on the VO$_2$ quality upon annealing with the STA method described in the manuscript. The grain size and surface roughness of three films (samples G-H-J) grown on PECVD, ALD, and thermal SiO$_2$ were measured by AFM (see Figure S3), and the quantitative values were extracted with the software Gwyddion.

**Table S2. VO₂ grains statistics of the samples measured in Figure S3.** Data analysis performed with Gwyddion (software version 2.59).

| STA treatment | Sample G | Sample H | Sample J |
|---|---|---|---|
| SiO$_2$ growth method | Thermal | PECVD | ALD |
| Average grain diameter size | 37 nm | 41 nm | 39 nm |
| RMS | 2.083 nm | 1.724 nm | 1.909 nm |

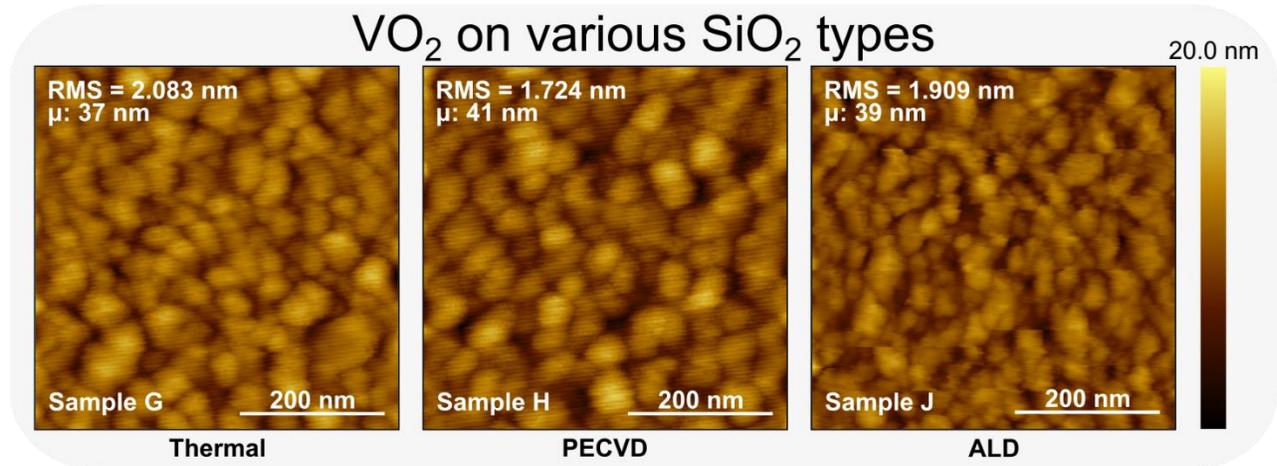

**Figure S3. AFM measurements of an STA treated VO₂ film on a SiO₂ substrate layer.** The substrate is grown (**a**) thermally (1 μm) (**b**) by PECVD (50 nm), or (**c**) ALD (50 nm). STA: Temp: 520 °C, O$_2$ pressure: 5 Pa, Annealing time: 5min.

On all three substrates, the obtained average grain size is around 40 nm, with a nominal variability of less than 4% between samples G, H, and J. The mean surface roughness values are also similar, showing a variation of about 20%.

In Figure S4, a 4-probe measurement of the resistivity against temperature of the $VO_2$ films grown on the PECVD and ALD $SiO_2$ layers reveals comparable hysteretic behaviors, with 1.5 orders of magnitude drop and a switching temperature close to 68 °C. These findings suggest that the method to grow $SiO_2$, with their corresponding differences in topography, purity and dangling bonds, has no or very little impact on the $VO_2$ structural properties and transition temperature. However, what stands out in Figure S4 is the remaining variability in the films resistivities measured before and after the phase-transition, similar to the behavior measured for the devices in Figure 6. Measuring the R-T characteristics of $VO_2$ grown on all three types of $SiO_2$ (ALD, PECVD, and Thermal) leads to nonoverlapping hysteresis curves, i.e. starting and ending at different resistivity values, as in Figures 1 and 2. Variability between $VO_2$ devices grown on $SiO_2$, regardless of the $SiO_2$ deposition technique, is inevitable. Hence, we investigated the growth of $VO_2$ grains on different substrate stack by adding an interlayer between the $VO_2$ and the $SiO_2$.

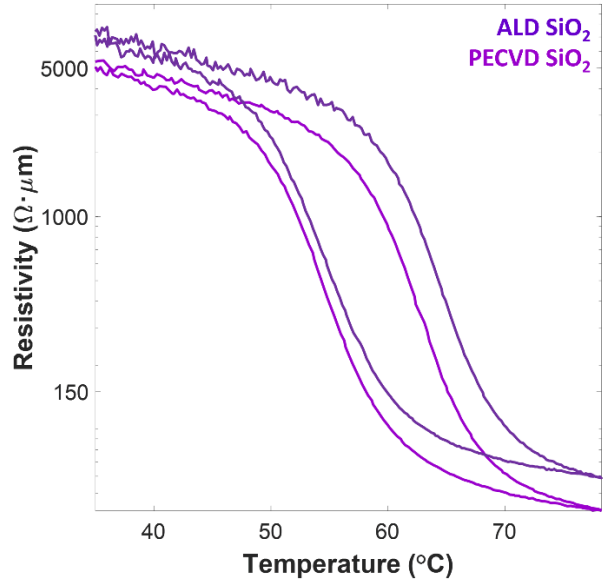

**Figure S4. R-T measurement of VO$_2$ samples grown on different SiO$_2$ types.** The VO$_2$ grains are annealed on 50 nm of PECVD or ALD deposited SiO$_2$ STA treated: Temp: 520 °C, O$_2$ pressure: 5 Pa, Annealing time: 5 min.

**Oscillation failure caused by high variability**

Figure S5a illustrates two VO$_2$-based oscillators operating independently without any coupling. The devices exhibit significant variability, evident in both the divergence in their fundamental oscillating frequencies (Figure S5b) and the notable difference in their insulator-to-metallic voltages, varying by more than 12%. When the oscillators are connected like in Figure 7d through a 330 pF capacitor, the expected outcome of stabilizing in an out-of-phase configuration does not occur; the oscillators simply ignore each other and do not lock in frequency (Figure S5d). The significant variation among the devices even makes 'strong coupling' unachievable (demonstrated with 1nF coupling in Figure S5d) since the oscillators continue to resist synchronization (Figure S5e).

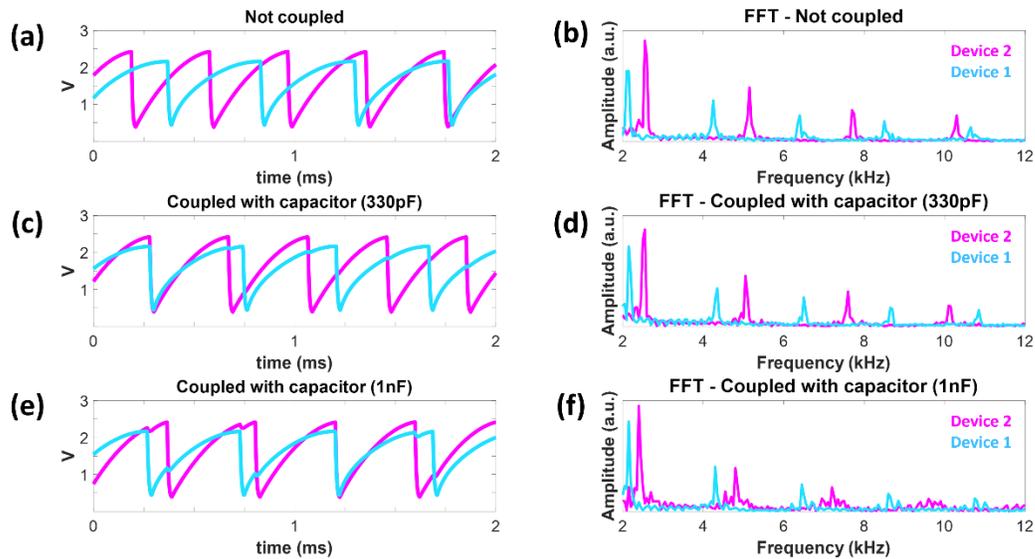

**Figure S5. Waveforms and Fast Fourier Transforms (FFT) of two VO$_2$-based oscillators coupled and uncoupled.** They show high variability when they are not coupled (**a** and **b**), capacitively (330 pF) coupled (**c** and **d**) and strongly capacitively (1nF) coupled (**e** and **f**). The three coupling schemes show how synchronization is impossible through a capacitor if the oscillators are too different to start with. $V_{DD}$ = 5.5 V. $R_s$ = 40 kΩ. $C_{ext}$ = 10 nF. Active area: 100 nm × 100 nm × 60 nm.

When connecting two oscillators with hitgh variability through a dissipative power connection, such as a resistive element, undesirable behaviors arise due to the exchange of current between their output nodes[2]. Figure S6a shows the waveforms of the oscillators when 'weakly coupled' through a 100 kΩ resistor. In this configuration, the network should stabilize in the in-phase configuration, contrary to what is shown in Figure S6a, although synchronization is achieved (Figure S6b). As the coupling resistance is reduced (Figure S6c), thereby increasing the coupling strength, current flows from one oscillator's output node to the other, sometimes preventing it from accumulating enough charge to trigger the phase-transition, as observed in Figure S6c. When the resistance is further decreased (Figure S6e), the phenomenon of 'oscillation death' occurs, meaning the exchange of current becomes too significant. This causes one oscillator to get trapped in the insulator state and the other in the metallic state. When the variability among devices is substantial, finding a balance to address all of these issues simultaneously becomes impossible.

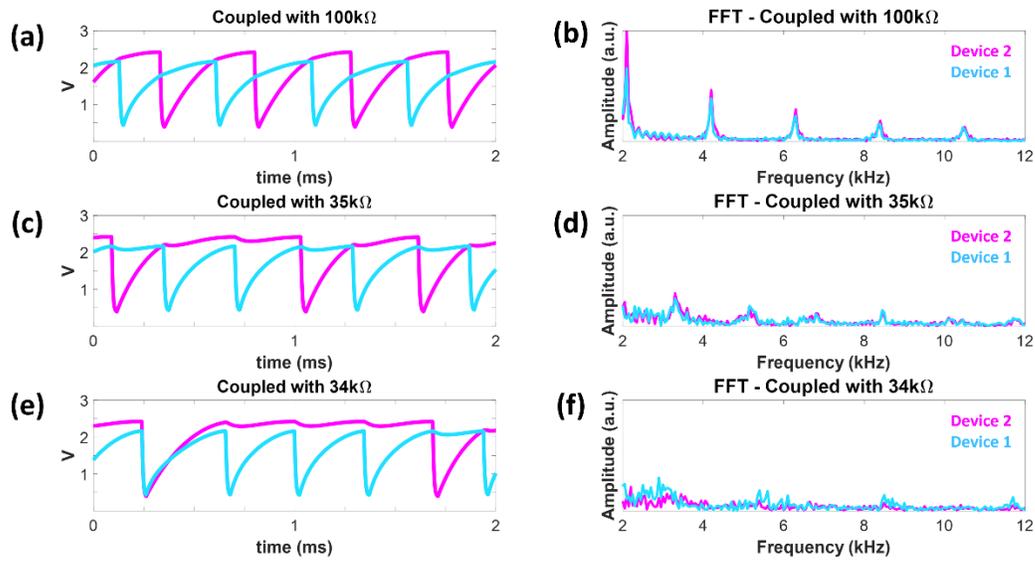

**Figure S6. Waveforms and Fast Fourier Transforms (FFT) of two VO₂-based oscillators showing high variability when they are weekly coupled.** The coupling is realized through a 100 kΩ resistor (**a** and **b**), and strongly coupled (35 kΩ and 34 kΩ, respectively) coupled (**c** and **d**, and **e** and **f**, respectively). (**a**) shows the unexpected out-of-phase synchronization in the case of weak coupling, while (**c**) and (**e**) show the progressive apparition of the 'oscillation death' phenomenon, which is impossible to balance with oscillators exhibiting high variability. $V_{DD}$ = 5.5 V. $R_s$ = 40 kΩ. $C_{ext}$ = 10 nF. Cross-section area: 100 nm × 100 nm × 60 nm.

## VO$_2$ Grown by RTA recipes and outcomes

**Table S3. RTA tests.** Rapid Thermal Annealing conditions tested to produce smooth granular VO$_2$ film on a Si/SiO$_2$ substrate. The best reproducible results are observed for the conditions: Temp: 470 °C, O$_2$ pressure: 5 Pa, Annealing time: 10 min.

| RTA Conditions | | | Film uniformity | Reproducibility | Presence of VO$_2$ |
|---|---|---|---|---|---|
| Final annealing temperature | O$_2$ pressure | Annealing time | | | |
| 430 °C | 5 Pa | 30 sec | Uniform | NO | NO |
| 520 °C | 5 Pa | 30 sec | Uniform | YES | NO |
| 470 °C | 25 Pa | 30 sec | Uniform | YES | NO |
| 470 °C | 25 Pa | 600 sec | Not uniform | NO | NO |
| 470 °C | 5 Pa | 450 sec | Uniform | NO | YES |
| **470 °C** | **5 Pa** | **600 sec** | **Uniform** | **YES** | **YES** |

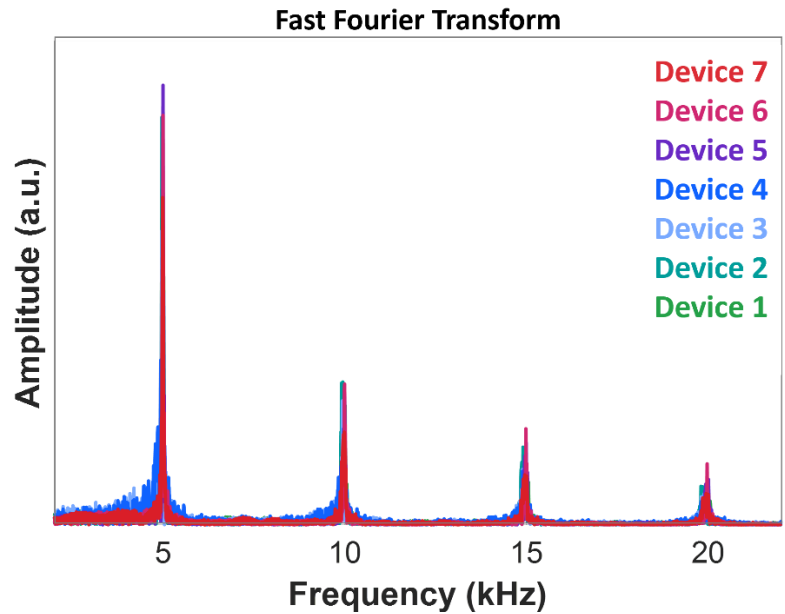

**Figure S7. Fast Fourier Transforms of the oscillatory response of 7 VO$_2$-based crossbar devices.** $V_{DD}$ = 5 V. $R_s \approx$ 40 kΩ. $C_{ext}$ = 10 nF.

**Accuracy of our XRR fitting**

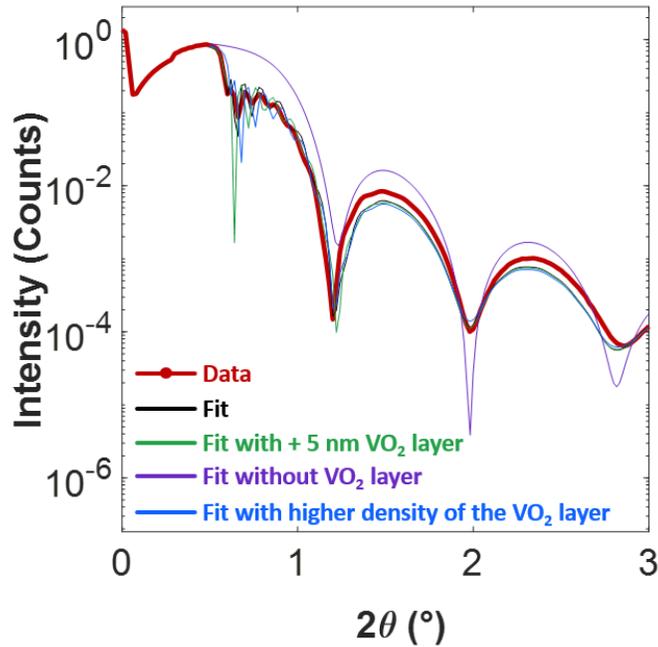

**Figure S8. Additional XRR fits**. The fits use multilayer models in which the $VO_2$ layer is removed (purple curve); the thickness of the $VO_2$ layer is increased of 5nm (green curve); the density of the $VO_2$ layer is increased to 4.4 g cm$^{-3}$ (blue curve). The original multilayer stack presented in the main manuscript gives the best agreement with the experimental data.